\documentclass{aa}  

\usepackage{graphicx}
\usepackage{capt-of}
\usepackage{txfonts}
\usepackage{lipsum}
\usepackage{lscape}             
\usepackage{placeins}
\usepackage{cuted}       
\usepackage[colorlinks=true,linkcolor=blue,citecolor=blue,urlcolor=blue]{hyperref}

\usepackage{todonotes}
\usepackage{xcolor}
\usepackage[normalem]{ulem}
\usepackage{cancel}

\AtBeginDocument{
  \nolinenumbers
  \global\let\linenumbers\relax
}

\begin{document}

\title{Evolution of superthin galaxies under Milgromian dynamics}

\author{
Zhe-Qi Huang\inst{1,2}\email{2024012028@bhu.edu.cn}
\and
Hao Chen\inst{3,4}\email{chenhao_qhnu@outlook.com}
\and
Xin-Lei Ge\inst{2}\email{gexl\_bhu@163.com}
\and
Cheng-Qun Pang\inst{1,5}\corrauth{xuehua45@163.com}
}

\institute{
School of Physics and Optoelectronic Engineering,
Ludong University, Yantai 264000, China
\and
College of Physical Science and Technology,
Bohai University, Jinzhou 121013, China
\and
College of Physics and Electronic Information Engineering,
Qinghai Normal University, Xining 810000, China
\and
Academy of Plateau Science and Sustainability,
Xining 810016, China
\and
Lanzhou Center for Theoretical Physics,
Key Laboratory of Quantum Theory and Applications of MoE,
and Key Laboratory of Theoretical Physics of Gansu Province,
Lanzhou University, Lanzhou, Gansu 730000, China
}

  \date{}
 
  \abstract  
   {Low-surface-brightness (LSB) superthin galaxies have extremely small vertical scale heights and low baryonic surface densities, but the physical mechanisms that allow such thin discs to form and survive over long timescales remain unclear.}
   {This work investigates the long-term evolution of the vertical structure of superthin galaxies within the framework of Milgromian dynamics (MOND). By constructing an observationally constrained model of UGC 7321, a typical superthin galaxy, we test whether its disc can maintain an extremely flattened structure in a MOND gravitational field. We also construct models with different values of the MOND depth index $D_{\rm M}$ to study how the global MOND depth affects disc evolution.} 
   {We perform three-dimensional hydrodynamical $N$-body simulations using the publicly available code Phantom of RAMSES. The evolution of the galactic disc is quantified using the ratio of the vertical scale height to the stellar-disc scale length $(h_z/R_{\rm D})$, together with Fourier amplitudes characterising non-axisymmetric structures and vertical buckling, measures of vertical heating, and the vertical restoring force.}
   {In the observationally constrained model of UGC 7321, the galaxy develops a strong bar and undergoes a buckling instability during the early stages of the simulation. The bar strength then decreases gradually, and the system eventually exhibits a weak bar structure. The vertical evolution reflects the combined effects of heating induced by non-axisymmetric structures and vertical confinement in the Milgromian potential. The stellar disc undergoes only limited vertical thickening, and the disc remains largely within the superthin regime, $h_z/R_{\rm D}<0.1$, after $5.0\,\mathrm{Gyr}$. The comparison of models with different $D_{\rm M}$ values suggests that models with lower $D_{\rm M}$ values, associated in our model suite with higher baryonic masses or more compact discs, exhibit stronger vertical heating and more significant disc thickening. By contrast, models with higher $D_{\rm M}$ values, corresponding to lower masses or more diffuse structures, tend to maintain a superthin structure.}
   {Overall, the simulation results indicate that superthin discs can remain vertically thin during long-term isolated evolution in MOND, and that the long-term maintenance of superthin structures is influenced, at least partly, by the degree to which a galaxy lies in the low-acceleration regime. To assess the generality of these findings, future work should extend the analysis to a broader sample of superthin galaxy models.}


   \maketitle
   \nolinenumbers

\section{Introduction}\label{sec:introduction}

Superthin galaxies are edge-on low-surface-brightness (LSB) disc galaxies without prominent bulges and with extremely flattened stellar discs~\citep{Goad1981,Karachentsev1993,deBlok1997,Matthews1999,Matthews2000,Kautsch2009}. 
Photometric studies of edge-on systems have shown that the degree of disc flattening varies among galaxy populations~\citep{Kregel2002,Bizyaev2017}.
Superthin galaxies are generally gas-rich, relatively unevolved late-type systems characterised by low star formation rates and low metallicities~\citep{Pickering1997,Abe1999,Uson2003,Mitronova2005,Matthews2008,Aditya2021}. 

In cosmological simulations within the standard $\Lambda$ cold dark matter ($\Lambda$CDM) framework, galaxies that naturally form and retain extremely thin discs over long timescales appear to be uncommon~\citep{Vogelsberger2014,Pillepich2018,Haslbauer2022,Hu2024}.
Internal sources of disc heating, such as bars, spiral arms, and giant molecular clouds, typically lead to disc thickening, unless the disc is stabilised by particular dark matter halo structures or favourable evolutionary histories~\citep{Saha2010,Aumer2016,Grand2016}. 
Studies of the vertical structure of superthin galaxies indicate that such systems are generally inferred to reside in dense, compact dark matter haloes, with halo core radii typically satisfying $R_{\rm C}/R_{\rm D}\lesssim2$, where $R_{\rm C}$ and $R_{\rm D}$ denote the halo core radius and the stellar-disc scale length, respectively. 

Such compact haloes may contribute to the long-term maintenance of superthin stellar discs~\citep{Khoperskov2010,OBrien2010,Banerjee2011,Banerjee2013,Banerjee2017,Kurapati2018,Sarkar2019}. 
In braneworld gravity, reproducing both the rotation curve and vertical scale height of superthin galaxies requires an additional gravitational contribution. This contribution is associated with the effective ``dark mass'' induced by the higher-dimensional Weyl stress~\citep{Komanduri2020}. 
Earlier photometric studies used the vertical-to-radial scale ratio to characterise intrinsic disc thickness~\citep{Reshetnikov2003,Bizyaev2004}. \citet{Banerjee2013} operationally defined a galaxy as superthin when the mean ratio within $R\leq3R_{\rm D}$ satisfies $\langle z_0/R_{\rm D}\rangle\leq0.1$. Here, $h_z$ is the scale parameter of the stellar vertical density profile, $\rho_{\rm star}\propto\operatorname{sech}^2(z/h_z)$, and corresponds to their $z_0$. We use $h_z/R_{\rm D}=0.1$ as an operational reference level, rather than a sharp dynamical boundary.

Recent collisionless $N$-body simulations~\citep{Aditya2025} further suggest that low surface brightness, and the correspondingly weak disc self-gravity, may help superthin discs to avoid thickening. The weak self-gravity of LSB discs can suppress bar formation and the associated heating, allowing initially superthin discs to maintain $h_z/R_{\rm D}<0.1$ during long-term isolated evolution, where $h_z$ is the stellar vertical scale height. In some simulations, higher stellar mass fractions and relatively shallow dark matter haloes are associated with stronger bar growth and more significant disc thickening~\citep{Aditya2023,Aditya2024,Aditya2025}. These explanations rely either on compact dark matter haloes within the Newtonian framework or on additional effective gravitational components in alternative-gravity models, together with specific structural conditions, evolutionary histories, or additional gravitational confinement.

Milgromian dynamics, originally introduced as modified Newtonian dynamics (MOND) by \citet{Milgrom1983}, provides an alternative framework for describing the mass discrepancies observed in galaxies. 
MOND postulates a fundamental acceleration scale, $a_0$, below which dynamics depart from their Newtonian behaviour. For systems with characteristic internal accelerations $g \ll a_0$, the dynamics approach the deep-MOND regime~\citep{Milgrom2009}. 
On galactic scales, a value of $a_0 \approx 1.2 \times 10^{-10}\,{\mathrm{m\,s^{-2}}}$ is commonly adopted on the basis of galaxy rotation-curve analyses~\citep{Begeman1991,Gentile2011}. 
In the deep-MOND regime, MOND predicts a mass--velocity scaling consistent with the baryonic Tully--Fisher relation~\citep{Milgrom1983,McGaugh2000}.
Large samples of galaxy rotation curves further reveal a tight radial acceleration relation between the observed centripetal acceleration and that predicted from the baryonic mass distribution~\citep{Lelli2016,McGaugh2016}. 

An action-based, non-relativistic modified-gravity formulation of MOND, known as the aquadratic Lagrangian theory (AQUAL), was developed by \citet{Bekenstein1984}. As a computationally convenient alternative to AQUAL, the quasi-linear formulation of MOND (QUMOND) was subsequently proposed by \citet{Milgrom2010}, in which the Newtonian potential is first calculated and then used to construct the source term of a second linear Poisson equation for the Milgromian potential. 
Earlier $N$-body studies investigated the local and global stability of galactic discs in MOND, as well as bar formation and evolution~\citep{Brada1999,Tiret2007,Tiret2008}. 
Furthermore, \citet{Lueghausen2015} later developed Phantom of RAMSES (POR), a Milgromian version of RAMSES that enables high-resolution $N$-body and hydrodynamical simulations of stellar and gaseous systems within the QUMOND framework~\citep{Nagesh2021}. 
Although embedding MOND in a complete cosmological framework remains challenging~\citep{Desmond2025}, the long-term vertical evolution considered here is primarily a galaxy-scale dynamical problem. It involves the low-acceleration regime characteristic of low-surface-brightness discs, where MOND has achieved some of its clearest empirical successes~\citep{Famaey2012,Desmond2025}. 
However, whether these successes extend to the long-term vertical evolution of superthin discs remains to be established.

In recent years, the POR framework has been applied to a variety of problems in galactic dynamics. \citet{Bilek2018} used it to investigate a past close flyby between the Milky Way (MW) and Andromeda (M31) and its possible role in the formation of the Local Group satellite planes. Subsequent three-dimensional hydrodynamical simulations suggested that an MW--M31 encounter with a pericentre distance of approximately $80\,\mathrm{kpc}$ could produce phase-space and orbital-pole distributions of tidal debris similar to those of the observed satellite planes~\citep{Banik2022}. 
POR has also been used to simulate the collapse of rotating gas clouds into exponential disc galaxies, as well as the evolution of isolated disc galaxies with star formation and stellar feedback included~\citep{Wittenburg2020,Nagesh2023}. 
In addition, MOND hydrodynamical simulations have been applied to specific galaxies to investigate the global stability and external-field response of M33, as well as the morphology and inferred inclination of the gas-rich ultra-diffuse galaxy AGC 114905~\citep{Banik2020,Banik20221}. 
Although previous studies have demonstrated the applicability of MOND simulations to a range of galaxy-scale dynamical problems, the long-term vertical evolution and structural survival of superthin discs have not yet been systematically investigated using hydrodynamical simulations within the MOND framework. 

UGC 7321 is a nearby, nearly edge-on, bulgeless Sd galaxy with a diffuse LSB stellar disc that is among the thinnest known, together with a gas-rich \ion{H}{i} disc and a well-measured rotation curve~\citep{Matthews1999,Matthews2000,Uson2003}. 
Within Newtonian mass models, its dynamics are consistent with a compact dark matter halo that dominates the gravitational field at essentially all radii~\citep{Banerjee2010}. Its relative isolation limits complications arising from strong environmental perturbations, while extensive optical, near-infrared, and \ion{H}{i} observations provide detailed constraints on its stellar and gaseous structure and kinematics~\citep{Matthews1999,Matthews2000,Uson2003}. Together, these properties make UGC 7321 a particularly suitable and well-observed benchmark for the present study. 
In this work, we focus on UGC 7321 and perform a series of three-dimensional hydrodynamical $N$-body simulations using POR. 
We construct an observationally constrained model of a superthin baryonic disc and test whether it can retain its superthin vertical structure during long-term isolated evolution within the QUMOND framework. We also examine how MOND depth is related to the long-term vertical evolution of superthin discs. 
Following the MOND depth index introduced by \citet{Eappen2026}, we use $D_{\rm M}$ as a global measure of a galaxy's MOND depth and construct a set of models with different values of $D_{\rm M}$. 
By comparing the evolution of these models over $5.0\,\mathrm{Gyr}$, we investigate whether variations in $D_{\rm M}$ are associated with systematic differences in vertical heating, the strength and evolution of non-axisymmetric structures, and late-time disc thickness. 

The remainder of this paper is organised as follows: Sect.~\ref{sec:initial} describes the initial conditions and simulation setup; Sect.~\ref{sec:results} presents the simulation results; and the main conclusions are summarised in Sect.~\ref{sec:conclusions}.

\section{Initial conditions and simulation setup}
\label{sec:initial}

We used the publicly available POR code to perform three-dimensional $N$-body and hydrodynamical simulations of UGC 7321. The simulations employ the QUMOND formulation, in which the Newtonian potential, $\Phi_{\rm N}$, is first obtained from the total baryonic mass density by solving the standard Newtonian Poisson equation. The resulting Newtonian field is then used to construct the source term of a second linear Poisson equation for the Milgromian gravitational potential, $\Phi$, given by
\begin{equation}
    \nabla^2 \Phi = \nabla \cdot \left[ \nu\left( \frac{|\nabla \Phi_{\rm N}|}{a_0} \right) \nabla \Phi_{\rm N} \right],  
\end{equation}
where $a_0$ is the MOND acceleration scale and $\nu$ is the QUMOND interpolating function. We adopted the ``simple'' form
\begin{equation}
    \nu(y) = \frac{1 + \sqrt{1 + 4/y}}{2},   
\end{equation}
where $y\equiv |\nabla\Phi_{\rm N}|/a_0$. This interpolating function approaches $\nu(y)\rightarrow 1$ in the Newtonian regime, $y\gg1$, and $\nu(y)\rightarrow y^{-1/2}$ in the deep-MOND regime, $y\ll1$.

UGC 7321 is modelled with separate exponential stellar and gas discs that have different radial scale lengths. The gas disc has the larger radial scale length, whereas the stellar disc is more compact. 
Both disc components are assumed to follow a $\operatorname{sech}^2$ vertical density profile, although their scale heights are specified separately. 
We adopt a stellar-disc scale length of $2.1\,\mathrm{kpc}$, consistent with previous observational and vertical-structure studies~\citep{Matthews1999,Sarkar2019}. 
To construct a model with a relatively high stellar mass while remaining consistent with the photometric constraints, we adopt the $B$-band colour--mass-to-light-ratio calibration of the infall galaxy-evolution model of \citet{Bell2001}, assuming their scaled Salpeter initial mass function (IMF). Specifically, the central colour, $B-R\approx1.2$, is used to estimate the stellar $B$-band mass-to-light ratio, $\Upsilon_B$, which is then combined with the central $B$-band luminosity surface density, $I_{B,0}=26.4\,L_\odot\,\mathrm{pc}^{-2}$, to determine the central stellar surface density, $\Sigma_0=\Upsilon_B I_{B,0}$~\citep{Banerjee2010}. For an exponential stellar disc, the total stellar mass is given by
\begin{equation}
M_{\rm star}=2\pi\Sigma_0R_{\rm D}^{2}.
\end{equation}

The observed \ion{H}{i} surface-density profile is not well described by a strictly single-exponential distribution. We therefore adopt an equivalent exponential scale length constrained jointly by the mass-weighted radial moment of the observed \ion{H}{i} surface-density profile and by the main and outer components of a double-Gaussian decomposition, obtaining a gas-disc scale length of $2.85\,\mathrm{kpc}$. 
We adopt the observed \ion{H}{i} mass of UGC 7321 reported by \citet{Uson2003} and apply a helium correction factor of 1.34 to obtain the total atomic-gas mass, including helium. 
Table~\ref{table1} lists the adopted parameters of the UGC 7321 model.
    
We used a MOND-adapted version of the Disk Initial Conditions Environment (DICE) code to generate the initial conditions corresponding to UGC 7321. 
The effective isothermal gas temperature was set to $T_{\rm gas}=2.5\times10^{4}\,\mathrm{K}$. 
This parameter sets the isothermal sound speed and hence the pressure support of the gas disc. It should not be interpreted as the literal thermodynamic temperature of a multiphase interstellar medium. 
The baryonic gas fraction, defined as $f_{\rm gas}=M_{\rm gas}/(M_{\rm star}+M_{\rm gas})$, was set to 0.484. 
\citet{Komanduri2020} adopted a vertical half-width at half-maximum (HWHM) of $0.105\,\mathrm{kpc}$ for the stellar disc of UGC 7321. For the vertical density profile adopted in this work, $\rho_{\rm star}(z)\propto\operatorname{sech}^{2}(z/h_z)$, the corresponding relation is $\mathrm{HWHM}=h_z\operatorname{arcosh}(\sqrt{2})\simeq0.8814\,h_z$. We adopt an initial stellar-disc scale height of $h_{z,0}=0.118\,\mathrm{kpc}$, within one per cent of the value inferred from the adopted HWHM. 
To impose a modest initial stability margin against local axisymmetric perturbations, we set the DICE parameter $Q_{\rm lim}$ to 1.25. 
The stellar disc was represented by $10^6$ equal-mass particles. 
We set the minimum and maximum refinement levels to levelmin=7 and levelmax=13, respectively, and adopted a cubic computational box with a side length of $512\,\mathrm{kpc}$. The finest 
cell size is determined by
$\Delta x_{\rm min}=L_{\rm box}/2^{\ell_{\rm max}}$, where
$L_{\rm box}=512\,\mathrm{kpc}$ and $\ell_{\rm max}=13$, yielding
$\Delta x_{\rm min}=62.5\,\mathrm{pc}$. The model was evolved for $5.0\,\mathrm{Gyr}$ to follow the long-term secular and vertical evolution of the disc. 

\begin{table}[t!]
\caption{Adopted parameters for the UGC 7321 model.}
\label{table1}
\centering
\renewcommand{\arraystretch}{1.18}
\setlength{\tabcolsep}{10pt}

\begin{tabular*}{0.65\columnwidth}
{@{\extracolsep{\fill}} c c}
\hline\hline

\quad Parameter & Value \\
\hline

\quad $\Sigma_{0}$\tablefootmark{a}
    & $56.3\,M_\odot\,\mathrm{pc}^{-2}$ \\

\quad $M_{\rm star}$\tablefootmark{b}
    & $1.6\times10^{9}\,M_\odot$ \\

\quad $M_{\rm gas}$\tablefootmark{c}
    & $1.5\times10^{9}\,M_\odot$ \\

\quad $R_{\rm D}$\tablefootmark{d}
    & $2.1\,\mathrm{kpc}$ \\

\quad $R_{\rm D,gas}$\tablefootmark{e}
    & $2.85\,\mathrm{kpc}$ \\

\hline
\hline
\end{tabular*}

\tablefoot{
\tablefoottext{a}{Central surface density of the stellar disc.}
\tablefoottext{b}{Total mass of the stellar disc.}
\tablefoottext{c}{Total atomic-gas mass, including the standard correction for helium.}
\tablefoottext{d}{Radial scale length of the stellar disc.}
\tablefoottext{e}{Radial scale length of the gas disc.}
}
\end{table}

\begin{figure}[!t]
   \centering
   \includegraphics[width=\hsize]{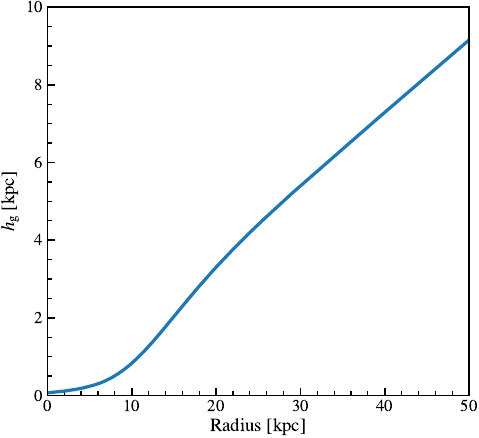}
   \caption{Radial profile of the initial gas-disc scale height, $h_{\rm g}(R)$, in the UGC 7321 model, assuming an effective isothermal gas temperature of $2.5\times10^{4}\,\mathrm{K}$.}
   \label{fig1}
\end{figure}

\begin{figure}[!t]
   \centering
   \includegraphics[width=\hsize]{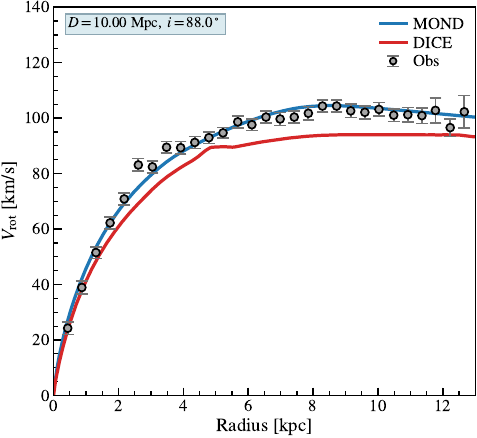}
   \caption{Rotation curve of UGC 7321. Grey points with error bars show the observed rotation-curve data. The blue curve shows the MOND rotation curve derived from the photometric mass model, assuming a fixed distance of $D=10.0\,\mathrm{Mpc}$ and an inclination of $i=88.0^\circ$, while the red curve shows the rotation curve of the DICE initial-condition model.}
   \label{fig2}
\end{figure}

The radial gas scale-height profile was constructed following the method of \citet{Banik2020}. For a vertically isothermal, self-gravitating thin gas disc with sound speed $c_s$ and vertical density distribution
$\rho_{\rm g}(R,z)\propto\operatorname{sech}^2[z/h_{\rm g}(R)]$, the gas scale height, $h_{\rm g}(R)$, satisfies 
\begin{equation}
    c_s^2 = \pi G h_{\rm g} \Sigma_{\rm g} = \frac{g_{\mathrm{N},z} h_{\rm g}}{2}, 
\end{equation}
where $\Sigma_{\rm g}$ is the gas surface density and
$g_{{\rm N},z}=2\pi G\Sigma_{\rm g}$ is the magnitude of the vertical Newtonian field generated by the gas disc in the thin-disc approximation. 

Following \citet{Banik2020}, we estimated the corresponding vertical-equilibrium condition in MOND using the algebraic MOND approximation, 
\begin{equation}
    c_s^2 \equiv \frac{\tilde{g}_{\mathrm{N},z} h_{\rm g}}{2} \nu\left( \frac{\sqrt{g_{\mathrm{N},R}^2 + \tilde{g}_{\mathrm{N},z}^2}}{a_0} \right), 
    \label{eq5}
\end{equation}
where $\tilde{g}_{\mathrm{N},z}$ is the magnitude of the total vertical Newtonian field generated by the combined stellar and gas discs, and $g_{\mathrm{N},R}$ is the magnitude of the radial Newtonian field generated by the total baryonic mass distribution. Because $\tilde{g}_{\mathrm{N},z}$ depends on the assumed vertical distribution of the gas, we solved Eq.~\eqref{eq5} iteratively for $h_{\rm g}(R)$ at each radius using the Newton--Raphson method. This procedure yielded the radial gas scale-height profile shown in Fig.~\ref{fig1}, which was used to initialise the gas disc in the hydrodynamical simulations. 

Figure~\ref{fig2} compares the observed rotation curve and its associated uncertainties reported by \citet{Uson2003} with two model predictions: a one-parameter algebraic MOND fit and the rotation curve of the DICE initial-condition model adopted in the simulations. 
The blue curve shows the one-parameter algebraic MOND fit. The observational inputs are adopted for $D=10.0\,\mathrm{Mpc}$ and $i=88^\circ$, and $a_0$ is held fixed. The gas contribution is calculated from the deprojected radial \ion{H}{i} surface-density profile of \citet{Uson2003}, normalised to the observed total \ion{H}{i} mass and corrected for helium. 
The stellar $B$-band mass-to-light ratio is the only fitted parameter. 
For the adopted luminosity, $L_B=1.1\times10^9\,L_\odot$, the fit yields a best-fitting mass-to-light ratio of $\Upsilon_B=2.99$. This value exceeds the range $\Upsilon_B\approx1.7$--$2.2$ predicted by the scaled Salpeter stellar-population models of \citet{Bell2001} at $B-R\approx1.2$, and implies a stellar mass approximately twice the independently inferred value listed in Table~\ref{table1}. By contrast, the red curve corresponds to the DICE initial-condition model, whose stellar and gas masses were fixed by independent photometric and \ion{H}{i} constraints. This model reproduces the overall radial trend of the observed rotation curve but underpredicts the rotation speed in the outer disc. We nevertheless retain this independently constrained, low-surface-density model because our primary aim is not to optimise the rotation-curve fit, but to construct an idealised superthin-disc model based on independent observational constraints and investigate whether it can retain its superthin vertical structure during long-term isolated evolution in MOND. 

\begin{figure}[!t]
   \centering
   \includegraphics[width=1.0\columnwidth]{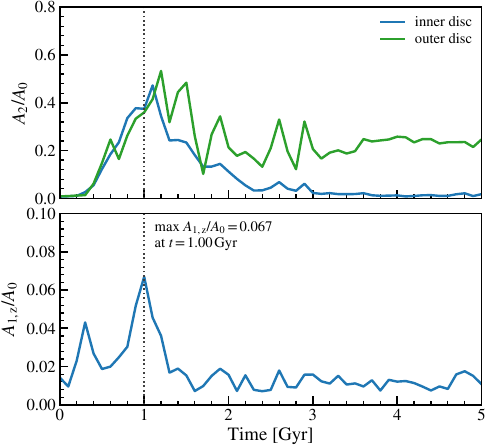}
   \caption{Time evolution of the normalised stellar Fourier amplitudes in the fiducial UGC 7321 model. The upper panel shows the $m=2$ amplitude, $A_2/A_0$, measured in the inner and outer disc regions, while the lower panel shows the vertical $m=1$ buckling amplitude, $A_{1,z}/A_0$. The vertical dotted line marks the time, $t=1.0\,\mathrm{Gyr}$, at which $A_{1,z}/A_0$ reaches its maximum.}
   \label{fig3}
\end{figure}

\section{Results and discussion}
\label{sec:results}
This section is organised in two stages. We first examine the observationally constrained fiducial model of UGC 7321 to determine whether an initially superthin stellar disc can retain its vertical structure during isolated evolution in MOND. We begin with the global morphological evolution and then quantify the growth of the bar, spiral structure, and bar buckling. We subsequently connect these non-axisymmetric features to the evolution of the stellar-disc scale height, vertical velocity dispersion, vertical heating, and Milgromian vertical restoring field, before examining the associated radial angular-momentum transport. We then extend the analysis to a sequence of models spanning different MOND depths. By comparing their morphologies, non-axisymmetric evolution, vertical heating, and final thicknesses, we assess which trends are associated with the global MOND depth. 

\subsection{Evolution of UGC 7321}
\label{subsec:ugc7321}

\begin{figure*}[!p]
    \centering
    \includegraphics[
        width=1.0\textwidth,
        height=1.0\textheight,
        keepaspectratio
    ]{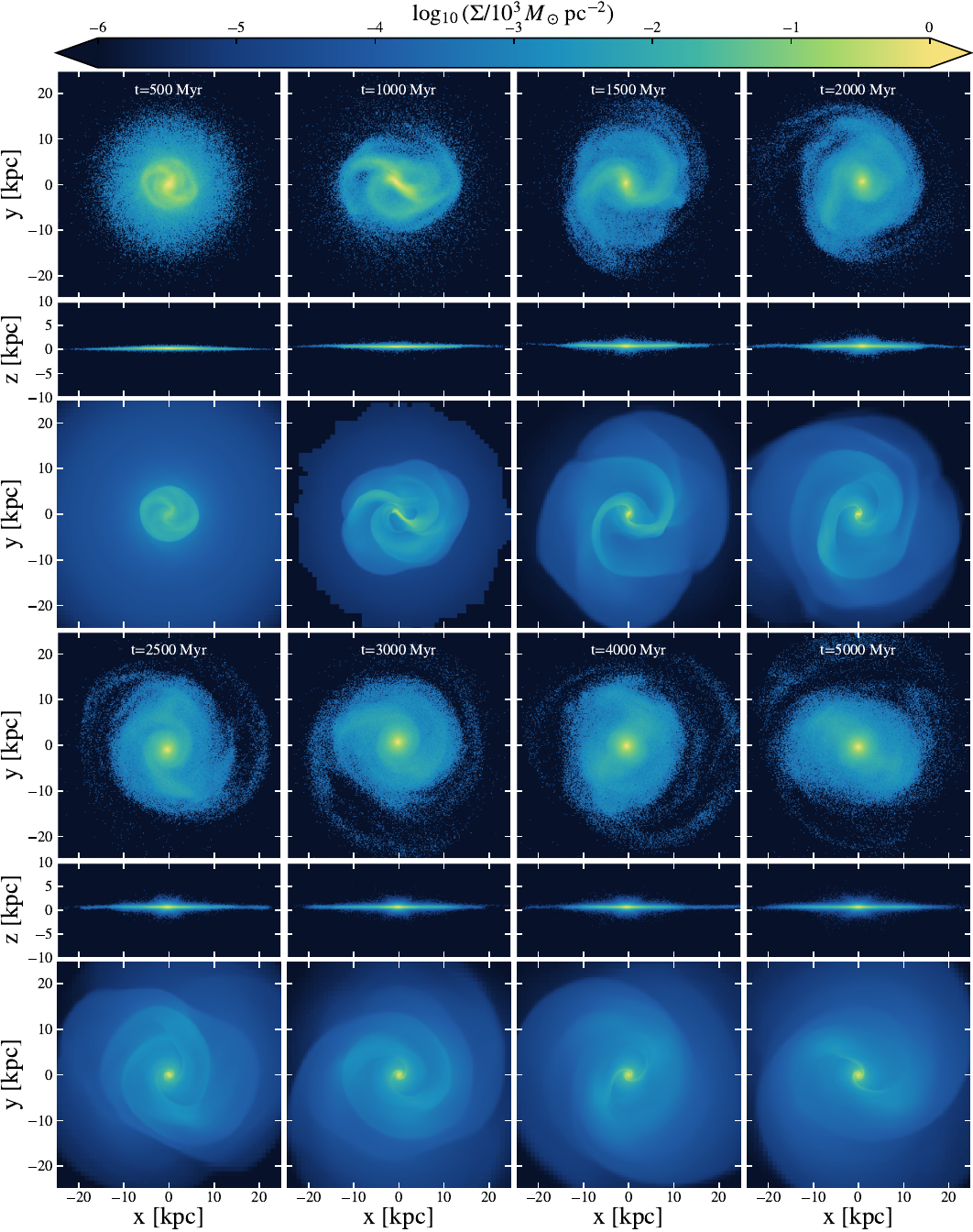}
    \caption{Evolution of the stellar and gas discs in the isolated UGC 7321 model. Within each three-row group, the upper, middle, and lower rows show the face-on stellar surface-density distribution, the corresponding edge-on stellar distribution, and the face-on gas surface-density distribution, respectively, at the evolutionary times indicated in the individual panels. The colour scale represents $\log_{10}\!\left[\Sigma/(10^3\,M_\odot\,\mathrm{pc}^{-2})\right]$.}
    \label{fig4}
\end{figure*}

Simulation snapshots were saved every $100\,\mathrm{Myr}$. Figure~\ref{fig4} presents selected face-on and edge-on stellar surface-density maps, together with the corresponding face-on gas surface-density maps. The snapshots are shown at $0.5\,\mathrm{Gyr}$ intervals from $0.5$ to $3.0\,\mathrm{Gyr}$, with additional snapshots at $4.0$ and $5.0\,\mathrm{Gyr}$. During the early evolution, a prominent bar develops in the inner stellar disc, while a two-armed spiral pattern forms in the outer disc.

\FloatBarrier
\clearpage
\clearpage

The gas disc exhibits a similar non-axisymmetric response, with prominent spiral structure developing over the same period. At later times, the inner stellar disc $m=2$ structure weakens, whereas non-axisymmetric features remain visible in the outer disc. 

\subsubsection{Non-axisymmetric structures}

We quantify the strength of the bar and spiral patterns in the simulated stellar disc using a Fourier decomposition~\citep{Saha2010,Saha2012,Saha2013,Saha2014}. 
For a given azimuthal mode $m$, the Fourier coefficients are calculated as 
\begin{equation}
a_m = \sum_i m_i \cos(m\phi_i), \quad b_m = \sum_i m_i \sin(m\phi_i), 
\end{equation}
where $m=0,1,2,\ldots$ is the azimuthal mode number, and $m_i$ and $\phi_i$ are the mass and azimuthal angle of the $i$th stellar particle, respectively. 
The corresponding normalised Fourier amplitude is defined as 

\begin{equation}
\frac{A_m}{A_0}
=
\frac{\sqrt{a_m^2+b_m^2}}
{\displaystyle\sum_{i}m_i}, 
\end{equation}
where $A_0=\sum_{i}m_i$ is the total stellar mass within the selected radial region. 

In this work, we focus on the $m=2$ mode and use $A_2/A_0$ to quantify the strength of the non-axisymmetric structure. The Fourier amplitudes are calculated separately for the inner and outer disc regions. The inner-disc amplitude primarily traces the bar, whereas the outer-disc amplitude primarily traces the two-armed spiral pattern. 

To quantify the bar and spiral structures separately, we determine the bar semi-major axis, $R_{\rm bar}$, from the radial behaviour of the $m=2$ Fourier phase. We then evaluate the Fourier amplitudes separately over the radial ranges $R<R_{\rm bar}$ and $R_{\rm bar}<R<4R_{\rm D}$, which represent the bar-dominated inner disc and the spiral-dominated outer disc, respectively. 

Similarly, we quantify the vertical asymmetry of the bar using the normalised vertical buckling amplitude, $A_{1,z}/A_0$, derived from the $m=1$ Fourier mode of the stellar distribution projected onto the $x$--$z$ plane~\citep{Li2024Bars}. At each simulation snapshot, the coordinate system is rotated so that the major axis of the bar coincides with the $x$-axis, while the disc rotation axis remains aligned with the $z$-axis. In this bar-aligned coordinate system, we select stellar particles within a region scaled to the bar size, following a prescription adapted from \citet{Li2023Buckling}: $|x|<R_{\rm bar}$, $|y|<0.25\,R_{\rm bar}$, and $|z|<0.4\,R_{\rm bar}$. The interval $-R_{\rm bar}<x<R_{\rm bar}$ is then divided into bins indexed by $k$ along the major axis of the bar. The normalised buckling amplitude is then calculated as 
\begin{equation}
\frac{A_{1,z}}{A_0}
=
\frac{
\displaystyle\sum_k
\sqrt{a_{1,k}^{\,2}+b_{1,k}^{\,2}}
}{
\displaystyle\sum_k a_{0,k}
},
\end{equation}
where $a_{1,k}$ and $b_{1,k}$ are the cosine and sine coefficients of the $m=1$ Fourier mode of the projected stellar distribution in the $k$th bin, respectively, and $a_{0,k}$ is the corresponding axisymmetric coefficient, equal to the total stellar mass in that bin. Thus, $\sum_k a_{0,k}$ is the total stellar mass within the selected bar region. 

The upper panel of Fig.~\ref{fig3} shows the evolution of the normalised stellar $m=2$ Fourier amplitude, $A_2/A_0$, in the inner and outer disc regions, while the lower panel shows the normalised vertical buckling amplitude, $A_{1,z}/A_0$. At $t=1.0\,\mathrm{Gyr}$, $A_{1,z}/A_0$ reached a maximum of approximately 0.067, indicating a pronounced episode of vertical asymmetry consistent with bar buckling. Shortly afterwards, at $t\approx1.1\,\mathrm{Gyr}$, the inner-disc $A_2/A_0$ reached a maximum of approximately 0.47, indicating that the bar was strongest at this time. The outer-disc $A_2/A_0$ reached a maximum of approximately 0.53 at $t\approx1.2\,\mathrm{Gyr}$, when the two-armed spiral pattern was most prominent. 

Following its early maximum, the inner-disc $m=2$ amplitude gradually declined. 
After $t\approx2.3\,\mathrm{Gyr}$, the inner-disc $A_2/A_0$ remained below 0.1. This indicates substantial weakening of the initially strong bar, with only a weak residual $m=2$ distortion in the central region. 
Similarly, after its peak, $A_{1,z}/A_0$ declined and subsequently remained below approximately 0.02, with no evidence for a second pronounced buckling episode. In contrast, after declining from its maximum, the outer-disc $A_2/A_0$ remained close to 0.3 for much of the subsequent evolution, consistent with the persistence of the two-armed spiral pattern. 

The early buckling episode was followed by the development of a weak boxy/peanut-shaped structure in the central region. This simulated morphology is qualitatively consistent with the photometric evidence for peanut-shaped outer isophotes in UGC 7321 reported by \citet{Pohlen2003} from deep $R$-band imaging. \citet{Pohlen2003} interpreted this feature as evidence for a large-scale stellar bar and a possible early stage of bar buckling. Our result demonstrates that a strong bar and a subsequent buckling episode can develop in an isolated MOND model constrained by the observed baryonic structure of UGC 7321. 
UGC 7321 is viewed nearly edge-on, with an inclination of
$i=88^\circ\pm1^\circ$~\citep{Uson2003}. The resulting line-of-sight
projection and superposition hinder the direct identification of its
in-plane non-axisymmetric structures. In particular, the
\ion{H}{i} data do not unambiguously distinguish a small bar from an
inner spiral arm~\citep{Uson2003}. 

\begin{figure}[!t]
   \centering
   \includegraphics[width=1.0\columnwidth]{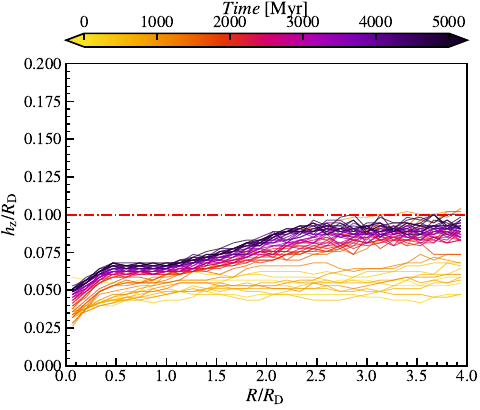}
   \caption{Radial profiles of the normalised stellar-disc scale height, $h_z/R_{\rm D}$, as a function of the normalised galactocentric radius, $R/R_{\rm D}$, at different simulation times. The colour of each curve indicates the simulation time, as shown by the colour bar above the panel. The red dash-dotted horizontal line marks the adopted superthin-disc threshold, $h_z/R_{\rm D}=0.1$.}
   \label{fig5}
\end{figure}

\begin{figure}[!t]
    \centering
    \includegraphics[width=1.0\columnwidth]{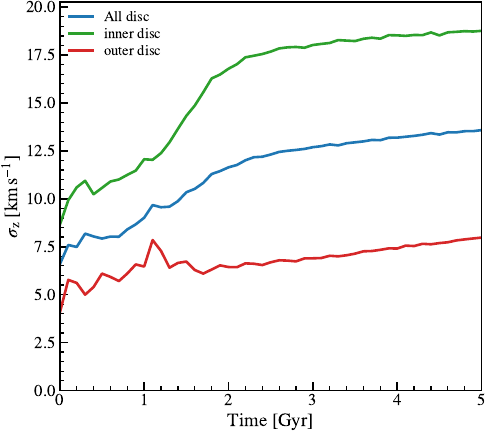}
    \vspace{0.35em}
    \includegraphics[width=1.0\columnwidth]{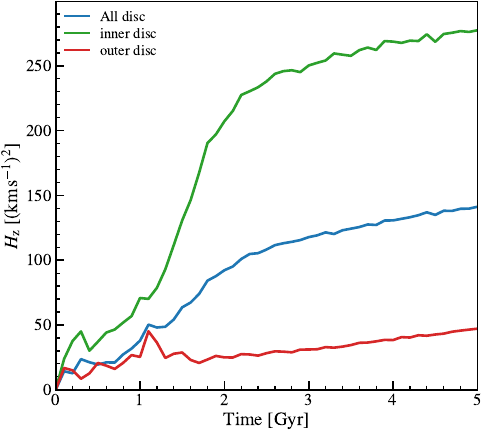}
    \caption{Time evolution of the vertical kinematics of the stellar disc in the UGC 7321 model. The upper panel shows the stellar vertical velocity dispersion, $\sigma_z$, while the lower panel shows the vertical heating parameter, $H_z(t)=\sigma_z^2(t)-\sigma_z^2(t_0)$, measured relative to the initial time $t_0$. The blue, green, and red curves correspond to the full, inner, and outer stellar disc regions, respectively.}
    \label{fig6}
\end{figure}

\begin{figure}[!t]
   \centering
   \includegraphics[width=1.0\columnwidth]{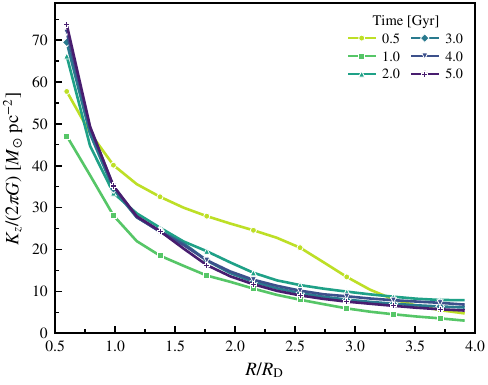}
   \caption{Radial profiles of the magnitude of the vertical restoring acceleration, expressed in surface-density units as $|K_z|/(2\pi G)$, evaluated at $|z'|=0.5\,\mathrm{kpc}$ in the UGC 7321 model. The profiles are shown as a function of the normalised galactocentric radius, $R/R_{\rm D}$, at different evolutionary times. The colours and symbols identify the simulation times, as indicated by the legend.}
   \label{fig7}
\end{figure}

We use the radial profiles of the normalised stellar-disc scale height, $h_z/R_{\rm D}$, at different evolutionary times to quantify changes in the disc thickness relative to its radial scale length. 
Throughout this work, $R_{\rm D}$ in both $h_z/R_{\rm D}$ and $R/R_{\rm D}$ denotes the initial stellar-disc scale length of the relevant model and is not re-fitted during the evolution. 
Here, $h_z(R,t)$ is defined through the vertical stellar density profile
$\rho_{\rm star}(R,z,t)=\rho_{\mathrm{star},0}(R,t)
\operatorname{sech}^2[z/h_z(R,t)]$.
Figure~\ref{fig5} shows the radial profiles of $h_z/R_{\rm D}$ at different evolutionary times over the range $R<4R_{\rm D}$. The model uses an initial scale-height ratio of $h_{z,0}/R_{\rm D}=0.056$, where $h_{z,0}$ denotes the prescribed scale parameter of the initial stellar vertical density profile. This ratio lies below the adopted operational superthin threshold of 0.1~\citep{Banerjee2013,Aditya2025}. Over $5.0\,\mathrm{Gyr}$ of isolated evolution, $h_z/R_{\rm D}$ increased across most of the disc, with more pronounced thickening at larger radii. By $t=5.0\,\mathrm{Gyr}$, the mean values of $h_z/R_{\rm D}$ in the adopted inner- and outer-disc regions had increased to approximately 1.5 and 1.8 times their initial values, respectively. Most of the stellar disc remained below the adopted threshold $h_z/R_{\rm D}=0.1$, although the late-time values at some outer radii approached or marginally exceeded this limit. 

Taken together, these results indicate that, although the UGC 7321 model experienced an early bar-buckling episode associated with the rapid growth of a strong bar, the buckling amplitude subsequently declined and no second pronounced buckling episode occurred. The simulated stellar disc therefore retains its overall superthin character over $5.0\,\mathrm{Gyr}$ of isolated evolution in MOND, despite moderate secular thickening, particularly in the outer disc. 

In the Newtonian simulations of \citet{Aditya2025}, weak disc self-gravity or a concentrated dark matter halo suppresses bar formation, limiting vertical thickening. More self-gravitating discs in shallow haloes instead develop bars and thicken. By contrast, our halo-free MOND model remains globally superthin after $5.0\,\mathrm{Gyr}$ despite strong bar formation and a possible early buckling episode. Although the two studies do not constitute a controlled comparison, our result suggests that bar suppression is not essential for superthin-disc survival in the present Milgromian model. The limited thickening is consistent with moderate vertical heating coexisting with confinement by the baryon-generated Milgromian restoring field. 

\subsubsection{Vertical heating and vertical force}

We characterise the vertical kinematic evolution of the stellar disc using the vertical velocity dispersion, $\sigma_z$, and quantify the associated vertical heating through its change relative to the initial state. The upper panel of Fig.~\ref{fig6} shows the time evolution of $\sigma_z$. For the full stellar disc, $\sigma_z$ increased overall from an initial value of approximately $6.5\,\mathrm{km\,s^{-1}}$ to approximately $13.5\,\mathrm{km\,s^{-1}}$ at $t=5.0\,\mathrm{Gyr}$. In the inner region, $R<R_{\rm D}$, $\sigma_z$ increased rapidly between approximately $1.0$ and $2.0\,\mathrm{Gyr}$ and subsequently approached a plateau. The timing of this increase is consistent with substantial vertical heating associated with the early growth and buckling of the bar. By contrast, in the outer-disc region, $2R_{\rm D}<R<4R_{\rm D}$, $\sigma_z$ increased by only approximately $4\,\mathrm{km\,s^{-1}}$, substantially less than the increase measured in the inner disc. This weaker increase is consistent with previous simulations showing that prominent spiral structure does not necessarily produce strong vertical heating~\citep{Grand2016}. 

Furthermore, we define the vertical heating parameter, $H_z(t)$, as the change in the squared vertical velocity dispersion relative to its initial value,   
\begin{equation}
H_z(t) = \sigma_z^2(t) - \sigma_z^2(t_0),   
\end{equation}
where $t_0=0$ denotes the initial simulation time. The lower panel of Fig.~\ref{fig6} shows the evolution of $H_z$, which serves as a measure of the cumulative vertical heating process within the stellar disc. 

To quantify the vertical restoring field near the stellar disc, we used the Osyris Python package to extract the Milgromian gravitational-acceleration field from the POR adaptive mesh. At each snapshot, the instantaneous disc centre, $\boldsymbol{x}_{\mathrm{c}}(t)$, and unit normal, $\hat{\boldsymbol n}(t)$, were determined from the stellar particle distribution. We then defined the disc-normal coordinate,
$z'=[\boldsymbol{x}-\boldsymbol{x}_{\mathrm{c}}(t)]
\cdot\hat{\boldsymbol n}(t)$,
and projected the gravitational acceleration along the disc normal,
$g_\perp(\boldsymbol{x},t)
=\boldsymbol{g}(\boldsymbol{x},t)
\cdot\hat{\boldsymbol n}(t)$.
The symmetrised restoring field was defined as 
\begin{equation}
K_z(R,t;z_0)
=
\frac{1}{2}
\left[
-\left\langle g_\perp\right\rangle_{\mathcal{V}_{+}(R)}
+
\left\langle g_\perp\right\rangle_{\mathcal{V}_{-}(R)}
\right],
\end{equation} 
where $R$ is the cylindrical radius measured in the instantaneous disc plane, and $\mathcal{V}_{\pm}(R)$ are the annular sampling volumes centred at $z'=\pm z_0$, each with a vertical half-width of $\Delta z$. The angle brackets denote volume-weighted averages over the selected adaptive mesh refinement (AMR) cells. We adopted $z_0=0.5\,\mathrm{kpc}$ and $\Delta z=0.1\,\mathrm{kpc}$, with each cell weighted by its volume, $\Delta V_i=(\Delta x_i)^3$. This symmetrised combination yields a scalar measure of the vertical restoring acceleration towards the instantaneous disc plane while cancelling acceleration components that act in the same direction on both sides of the disc. We divided the radial range $0.5\leq R/R_{\rm D}<4$ into 18 annuli and sampled the restoring field at intervals of $0.5\,\mathrm{Gyr}$. Figure~\ref{fig7} presents the resulting radial profiles in terms of $|K_z|/(2\pi G)$. This quantity has the dimensions of surface density but should not be interpreted as a direct measurement of the baryonic surface density of the disc. 

At all sampled times, $|K_z|$ generally decreases with radius, indicating that the inner disc is consistently subject to stronger vertical gravitational confinement. At $t=1.0\,\mathrm{Gyr}$, $|K_z|$ is approximately $18$--$61\%$ lower than at $t=0.5\,\mathrm{Gyr}$ across the sampled radial range. The subsequent evolution exhibits clear radial variations. Relative to the profile at $t=0.5\,\mathrm{Gyr}$, the value at $t=5.0\,\mathrm{Gyr}$ is approximately $28\%$ higher at $R\simeq0.60\,R_{\rm D}$, approximately $48$--$56\%$ lower over $R\simeq2.0$--$2.5\,R_{\rm D}$, and approximately $14\%$ higher at $R\simeq3.9\,R_{\rm D}$. These differences indicate that the vertical restoring field evolves in a radius-dependent manner as the mass distribution changes, rather than undergoing a simple overall rescaling with time. 

Taken together with the modest increase in disc thickness and the comparatively weak vertical heating of the outer disc, the continued presence of a substantial Milgromian restoring field is consistent with a role in limiting vertical expansion. Within the present model, the balance between vertical heating and gravitational confinement provides a plausible explanation for why the stellar disc remains globally superthin after $5.0\,\mathrm{Gyr}$. 

\subsubsection{Radial angular-momentum transport}

\begin{figure*}[!t]
    \centering
    \includegraphics[width=\textwidth]{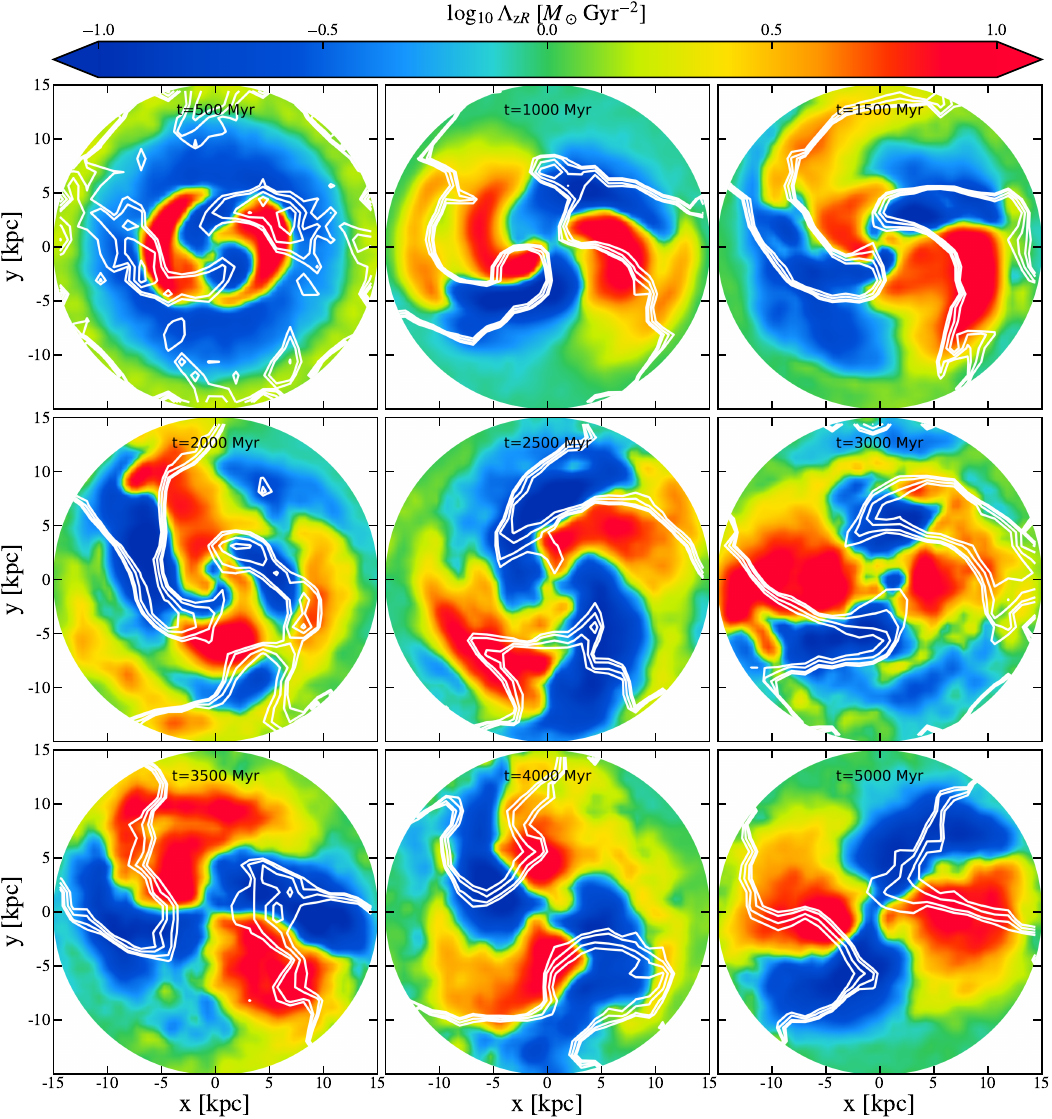}
    \caption{Face-on maps of the radial flux of the $z$-component of angular momentum, $\Lambda_{zR}$, in the UGC 7321 model at the evolutionary times indicated in each map. Positive and negative values represent outward and inward radial transport of angular momentum, respectively. The colour scale shows the transformed flux values used for visualisation, while the white contours indicate selected levels of the stellar surface-density distribution.}
    \label{fig8}
\end{figure*}

We calculate the radial flux of the $z$-component of angular momentum to diagnose transport and redistribution within the disc~\citep{Grand2016}. This diagnostic allows us to investigate their possible connection with vertical evolution. 

The angular-momentum flux tensor is defined as 
\begin{equation}
\Lambda_{i\alpha}
=
\epsilon_{ijk}x_j\rho_{\rm star}
\overline{v_k v_\alpha},
\end{equation}
where $\epsilon_{ijk}$ is the Levi--Civita symbol, $x_j$ is the
$j$th Cartesian component of the position vector, $\rho_{\rm star}$ is
the local stellar mass density, and $v_k$ and $v_\alpha$ are the
velocity components in the $k$ and $\alpha$ directions,
respectively. The overbar denotes a local average over the stellar
particles within each spatial bin. 

For the radial transport of the $z$-component of angular momentum,
we set $i=z$ and $\alpha=R$, yielding 
\begin{equation}
\Lambda_{zR}
=
R\rho_{\rm star}\overline{v_\phi v_R},
\end{equation}
where $R=\sqrt{x^2+y^2}$ is the cylindrical radius, and $v_R$ and $v_\phi$ are the radial and azimuthal velocity components, respectively. Positive $\Lambda_{zR}$ indicates the outward radial transport of the $z$-component of angular momentum, while negative values indicate inward transport.

Figure~\ref{fig8} presents face-on maps of the radial angular-momentum flux, $\Lambda_{zR}$, at $0.5\,\mathrm{Gyr}$ intervals from $0.5$ to $4.0\,\mathrm{Gyr}$, together with an additional map at $5.0\,\mathrm{Gyr}$. At $t=0.5\,\mathrm{Gyr}$, two regions of positive flux were approximately aligned with the early two-armed stellar pattern. Regions of negative flux were also present between the arms and at larger radii, indicating simultaneous inward angular-momentum transport. These alternating positive and negative-flux regions show that radial angular-momentum transport and redistribution were already active at this early stage. 

At $t=1.0\,\mathrm{Gyr}$, the $\Lambda_{zR}$ map showed the clearest central quadrupolar pattern, with alternating regions of outward and inward angular-momentum flux. This morphology is consistent with bar-driven radial motions and coincides with the epoch when the $m=2$ mode was strongest. By $t=2.0\,\mathrm{Gyr}$, the angular-momentum flux pattern extended into the intermediate and outer disc and developed a more spiral-like morphology. As the inner-disc $m=2$ amplitude declined, the central quadrupolar pattern became less prominent. At later times, regions of both outward and inward flux persisted, but became more diffuse and were most evident at larger radii. 

\begin{table*}[ht!]
\caption{Baryonic masses and structural parameters of the fiducial UGC 7321 model M1 and the comparison models M2--M5.} 
\label{table2}     
\centering                                                   
\begin{tabular}{l c c c c c c c c}            
\hline\hline              
Model & $M_{\rm bar}$ ($10^{9}\,M_\odot$) & $M_{\rm star}$ ($10^{9}\,M_\odot$) & $M_{\rm gas}$ ($10^{9}\,M_\odot$) & $f_{\rm gas}$ & $R_{\rm D,gas}$ $(\mathrm{kpc})$ & $R_{\rm D}$ $(\mathrm{kpc})$ & $D_{\rm M}$ & $T_{\rm gas}$ $(\mathrm{K})$ \\ 
\hline                      
M1 & 3.10 & 1.60 & 1.50 & 0.484 & 2.850 & 2.100 & 0.81 & $2.5\times10^{4}$ \\
M2 & 6.20 & 3.20 & 3.00 & 0.484 & 2.850 & 2.100 & 0.69 & $2.5\times10^{4}$ \\
M3 & 3.10 & 1.60 & 1.50 & 0.484 & 1.995 & 1.470 & 0.69 & $2.5\times10^{4}$ \\
M4 & 1.55 & 0.80 & 0.75 & 0.484 & 2.850 & 2.100 & 0.89 & $2.5\times10^{4}$ \\
M5 & 3.10 & 1.60 & 1.50 & 0.484 & 4.275 & 3.150 & 0.89 & $2.5\times10^{4}$ \\
\hline
\hline
\end{tabular}
\tablefoot{The columns list (1) the model name; (2) the total baryonic mass, $M_{\rm bar}=M_{\rm star}+M_{\rm gas}$; (3) the stellar mass; (4) the gas mass; (5) the gas fraction, $f_{\rm gas}=M_{\rm gas}/M_{\rm bar}$; (6) the radial scale length of the gas disc, $R_{\rm D,gas}$; (7) the radial scale length of the stellar disc, $R_{\rm D}$; (8) the MOND depth index, $D_{\rm M}$; and (9) the initial effective isothermal gas temperature, $T_{\rm gas}$. Model M1 is the fiducial UGC 7321 model, whereas models M2--M5 were constructed by varying the baryonic masses or disc scale lengths, as described in the text.}
\end{table*}

\begin{figure*}[t!]
    \centering
    \includegraphics[width=\textwidth]{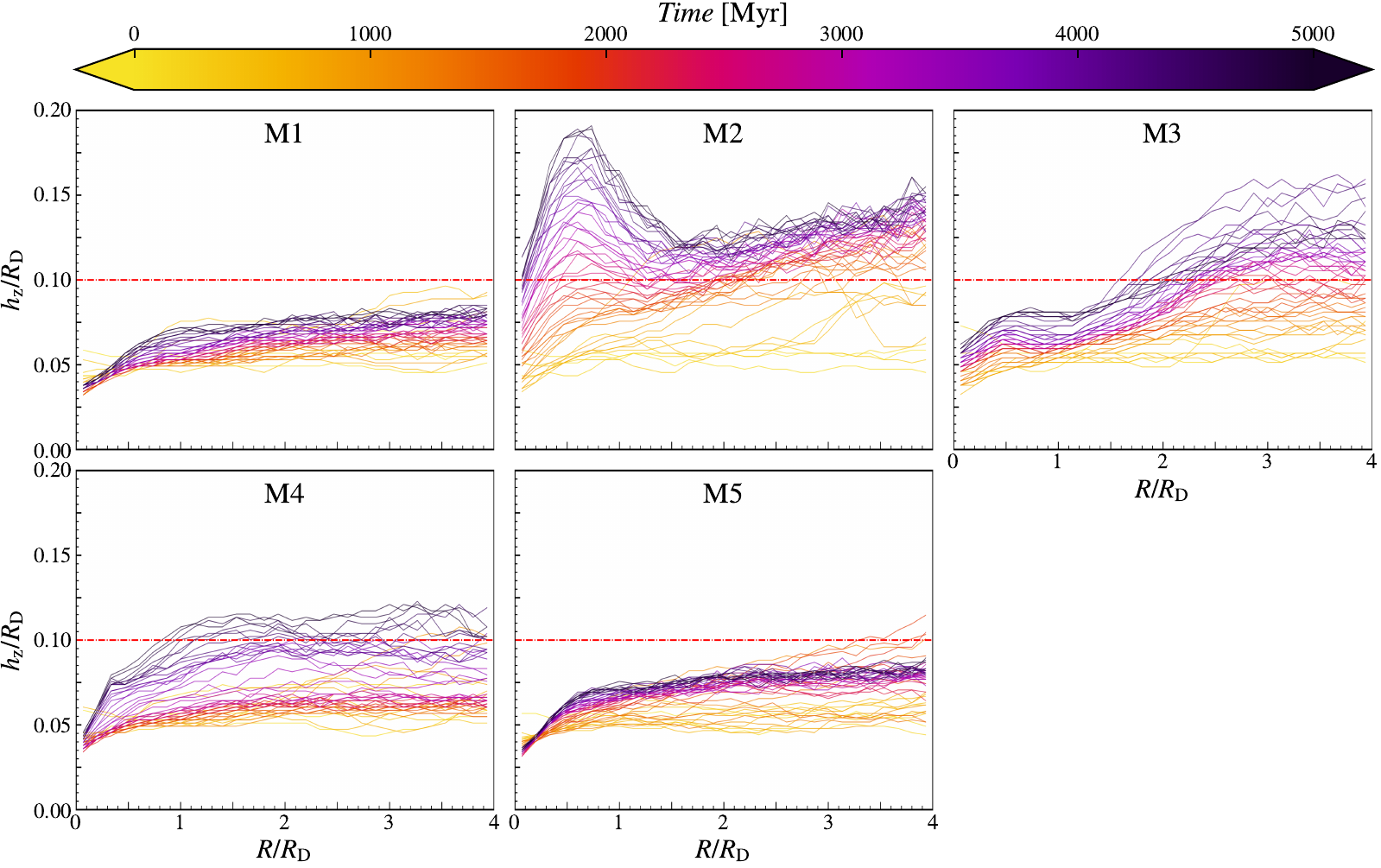}
    \caption{Radial profiles of the normalised stellar-disc scale height, $h_z/R_{\rm D}$, as a function of the normalised radius, $R/R_{\rm D}$, for models M1--M5. The panels are labelled by model name, and $R_{\rm D}$ denotes the stellar-disc scale length of the corresponding model. The colour of each curve indicates the simulation time, as shown by the upper colour bar. The red dash-dotted horizontal line marks the adopted superthin-disc threshold, $h_z/R_{\rm D}=0.1$.}
    \label{fig9}
\end{figure*}

These results suggest that substantial angular-momentum redistribution occurred within the disc over the simulated evolution. However, the comparatively weak vertical heating, particularly in the outer disc, suggests that much of this redistribution remained within the disc plane rather than being converted into vertical random motion. 

\subsection{Dependence of vertical evolution on MOND depth}

\subsubsection{Definition of the MOND depth index}
\citet{Eappen2026} introduced the dimensionless MOND depth index, $D_{\rm M}$, to quantify the fraction of a system's baryonic mass lying beyond its characteristic MOND radius. The characteristic MOND radius, $r_{\rm M}$, is defined by $GM_{\rm bar}/r_{\rm M}^{2}=a_0$, yielding 
\begin{equation}
r_{\rm M} = \sqrt{\frac{G M_{\rm bar}}{a_0}}. 
\end{equation} 

Here, $G$ is the gravitational constant, $M_{\rm bar}$ is the total baryonic mass of the system, and $a_0$ is the MOND acceleration scale. This radius provides a convenient boundary for separating the baryonic mass lying inside and outside the characteristic MOND transition scale. Baryonic matter lying beyond $r_{\rm M}$ is associated with the low-acceleration domain in which MOND effects become increasingly important. 

On this basis, the MOND depth index is defined as 
\begin{equation}
D_{\rm M}
=
1-
\frac{M_{\rm bar}(<r_{\rm M})}
     {M_{\rm bar}}
=
\frac{M_{\rm bar}(>r_{\rm M})}
     {M_{\rm bar}}.
\end{equation} 

Here, $M_{\rm bar}(<r_{\rm M})$ denotes the baryonic mass enclosed within $r_{\rm M}$, so that $D_{\rm M}$ directly measures the fraction of the total baryonic mass lying outside this radius. By construction, $D_{\rm M}\rightarrow0$ when most of the baryonic mass lies inside $r_{\rm M}$, as expected for a relatively compact, predominantly high-acceleration system. Conversely, $D_{\rm M}\rightarrow1$ when most of the baryonic mass lies outside $r_{\rm M}$, corresponding to a more diffuse system whose baryonic distribution extends predominantly into the low-acceleration MOND regime. 

The fraction of baryonic mass beyond the MOND radius reflects how far the baryonic distribution extends into the low-acceleration MOND regime. The index thus provides a global measure of MOND depth for galaxies with different structures. 

We adopt model M1 as the fiducial model for this comparison. Its baryonic masses and disc scale lengths are constrained by the adopted observational properties of UGC 7321 and are listed in Table~\ref{table2}. Model M1 has the same physical mass model and initial baryonic structure as the high-resolution fiducial UGC 7321 model described in Sect.~\ref{subsec:ugc7321}, but it is evolved at a lower numerical resolution. 
All five comparison models were initialised with the same prescribed scale-height ratio, $h_{z,0}/R_{\rm D}=0.056$, where $h_{z,0}$ denotes the prescribed scale parameter of the initial stellar vertical density profile. Consequently, $h_{z,0}$ scales with the adopted initial stellar-disc scale length and equals $0.118\,\mathrm{kpc}$ for M1, M2, and M4, $0.082\,\mathrm{kpc}$ for M3, and $0.176\,\mathrm{kpc}$ for M5. This design gives the models the same input relative thickness and allows their subsequent values of $h_z/R_{\rm D}$ to be compared directly. 
For each model, $D_{\rm M}$ is calculated from the cumulative baryonic mass of the combined stellar and gas discs enclosed within $r_{\rm M}$. We construct five models, M1--M5, spanning different values of $D_{\rm M}$ by varying the baryonic masses or disc scale lengths. 

Relative to M1, the stellar and gas masses were doubled in M2 and halved in M4, while the disc scale lengths were kept unchanged. 
For M3 and M5, the baryonic masses were unchanged from M1, while both disc scale lengths were multiplied by 0.7 and 1.5, respectively. 
These choices deliberately produce two matched-depth pairs through different structural changes: M2 and M3 both have $D_{\rm M}=0.69$, whereas M4 and M5 both have $D_{\rm M}=0.89$. The pairs therefore test whether a common global MOND depth implies a common evolutionary outcome when it is obtained by changing the baryonic mass or the disc scale lengths. 
All five models were evolved with levelmax=12, corresponding to a finest cell size of $125\,\mathrm{pc}$ for the adopted $512\,\mathrm{kpc}$ computational box. These simulations are intended primarily to compare the relative vertical evolution of models with different values of $D_{\rm M}$ and to capture their broad evolutionary trends rather than detailed small-scale structure. Model M1 therefore provides a lower-resolution counterpart to the high-resolution fiducial simulation. 

\subsubsection{Comparison of models with different MOND depths}

\begin{figure}[t!]
    \centering
    \includegraphics[width=1.0\columnwidth]{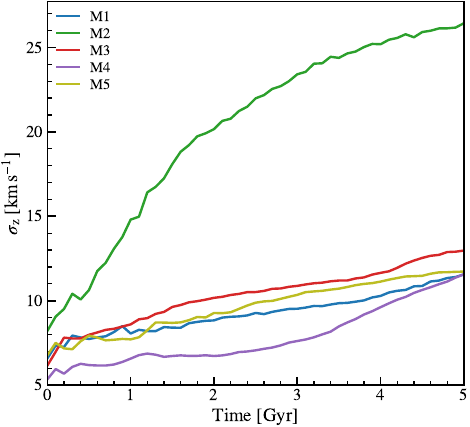}
    \vspace{0.35em}
    \includegraphics[width=1.0\columnwidth]{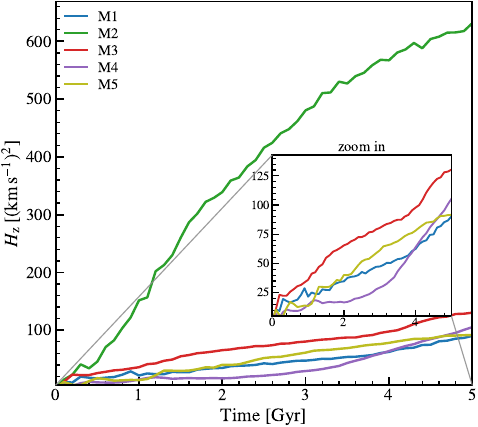}
    \caption{Time evolution of the vertical kinematics of models M1--M5. The upper panel shows the vertical velocity dispersion, $\sigma_z$, while the lower panel shows the vertical heating parameter, $H_z$. The inset in the lower panel provides a zoomed-in view of the $H_z$ curves to highlight the differences among the models.}
    \label{fig10}
\end{figure}

We evolved all five models for $5.0\,\mathrm{Gyr}$. Figure~\ref{fig11} presents face-on and edge-on views of the stellar distributions at $t=5.0\,\mathrm{Gyr}$, while Fig.~\ref{fig14} shows the time evolution of the normalised $m=2$ amplitude, $A_2/A_0$, and the normalised buckling amplitude, $A_{1,z}/A_0$, for each model. The five models developed distinct long-term morphologies and vertical evolutionary histories. 

To assess the sensitivity of the results to numerical resolution, we compare the high-resolution fiducial model described in Sect.~\ref{subsec:ugc7321} with model M1, which has the same physical parameters except for the maximum refinement level, levelmax. Both simulations exhibit rapid early growth of the $m=2$ non-axisymmetric structure and a brief enhancement of vertical asymmetry, although the strengths and occurrence times of the corresponding peaks are not identical. In the higher-resolution model, the inner- and outer-disc $A_2/A_0$ reach approximately 0.47 and 0.53 at $t=1.1$ and $1.2\,\mathrm{Gyr}$, respectively, while $A_{1,z}/A_0$ reaches approximately 0.067 at $t=1.0\,\mathrm{Gyr}$. By comparison, the inner-disc $A_2/A_0$ of M1 reaches approximately 0.41 at $t=0.9\,\mathrm{Gyr}$, and the peak in vertical asymmetry occurs earlier, reaching approximately 0.076 at $t=0.5\,\mathrm{Gyr}$. Meanwhile, the outer-disc $A_2/A_0$ continues to exhibit strong fluctuations at later times and reaches approximately 0.46 at $t=4.0\,\mathrm{Gyr}$. Therefore, the precise peak values, phases, and late-time strengths of the non-axisymmetric modes are relatively sensitive to spatial resolution. 

The final face-on view of M1 retains more prominent non-axisymmetric structures in the central and outer disc than the higher-resolution model, consistent with its larger late-time $A_2/A_0$. 
Figure~\ref{fig12} directly compares the final radial profiles of $h_z/R_{\rm D}$ in the high-resolution fiducial model and M1 at $t=5.0\,\mathrm{Gyr}$. The radial-bin mean over $0\leq R/R_{\rm D}<4$ is approximately 0.081 in the high-resolution model and 0.072 in M1, with radial ranges of approximately 0.049--0.100 and 0.038--0.083, respectively. The lower-resolution mean is therefore approximately 11\% smaller. The profiles nearly overlap over $0.5\lesssim R/R_{\rm D}\lesssim1.2$, where the mean absolute difference between corresponding bins is approximately 0.003. Beyond $R/R_{\rm D}\simeq1.3$, the high-resolution profile is larger, with a separation of approximately 0.02 near $4R_{\rm D}$. An offset is also present in the innermost bins. Nevertheless, the high-resolution model remains below the adopted threshold over most of the sampled disc and approaches it at the outer edge, while M1 remains below it throughout. The two runs therefore agree on the qualitative persistence of a very thin disc. 

At late times, M2 showed neither a strong bar nor prominent spiral arms. It underwent a pronounced early buckling episode, and its edge-on morphology at $t=5.0\,\mathrm{Gyr}$ was visibly thicker than that of M1, consistent with the scale-height measurements shown in Fig.~\ref{fig9}. M3 likewise showed no strong late-time bar or prominent spiral pattern, but retained a weak central thickening. Its buckling amplitude remained elevated for a longer period than in the other models. The face-on view of M4 showed an inner bar and prominent outer spiral structure, while its $A_{1,z}/A_0$ evolution indicated a later but weaker buckling episode. The larger disc scale lengths adopted for M5 produced a more diffuse baryonic distribution. At late times, M5 retained weak central thickening together with outer spiral features, while its $A_{1,z}/A_0$ evolution showed only a single pronounced early buckling episode. The contrasting evolution of M2 and M3, as well as that of M4 and M5, demonstrates that models with the same $D_{\rm M}$ can develop different morphologies when their masses and scale lengths differ. 

Figure~\ref{fig9} shows the radial profiles of $h_z/R_{\rm D}$ at different evolutionary times over the range $R<4R_{\rm D}$ for each model. Models M1 and M5 remained predominantly below the adopted superthin-disc threshold throughout the simulation. In M4, $h_z/R_{\rm D}$ exceeded the adopted threshold of 0.1 over most of the sampled radial range at late times. M2 exhibited the strongest vertical thickening among the five models and exceeded the superthin-disc threshold over a substantial radial range. In M3, the inner-disc value of $h_z/R_{\rm D}$ remained low during the early evolution, whereas the outer-disc scale height increased progressively after $t\approx2.0\,\mathrm{Gyr}$ and later exceeded 0.1. 

Figure~\ref{fig10} compares the evolution of $\sigma_z$ and $H_z$ among the five models. M2 exhibited the strongest vertical heating, with $\sigma_z$ increasing overall from approximately $8\,\mathrm{km\,s^{-1}}$ initially to approximately $26\,\mathrm{km\,s^{-1}}$ at $t=5.0\,\mathrm{Gyr}$. In M3, $\sigma_z$ increased steadily throughout the evolution, and both $\sigma_z$ and $H_z$ indicated that it experienced the second-strongest vertical heating. By contrast, M1, M4, and M5 underwent substantially weaker vertical heating. 

Figure~\ref{fig13} compares the final radial profiles of $h_z/R_{\rm D}$ for models M1--M5 at $t=5.0\,\mathrm{Gyr}$. Table~\ref{table3} lists the mean ($\langle h_z/R_{\rm D}\rangle_R$), minimum, and maximum values of $h_z/R_{\rm D}$ across the sampled radii, together with
the final $H_z$ for each model. M2 has the largest mean, $\langle h_z/R_{\rm D}\rangle_R=0.142$, and the largest $H_z=631\,(\mathrm{km\,s^{-1}})^2$. Its minimum and maximum values of $h_z/R_{\rm D}$ are 0.102 and 0.189, respectively, indicating that the profile remains above the adopted threshold throughout the sampled disc. Although M3 has the same $D_{\rm M}$ as M2, its mean and $H_z$ are smaller, at 0.100 and $130\,(\mathrm{km\,s^{-1}})^2$, respectively. Its minimum and maximum values of $h_z/R_{\rm D}$ are 0.059 and 0.135, respectively. The profile begins to exceed the adopted threshold at approximately $R=2.2R_{\rm D}$ and remains above it at larger radii. M1 and M5 have the smallest means, 0.072 and 0.074, respectively. Their maximum values, 0.083 and 0.089, respectively, remain below the adopted threshold. Models M4 and M5, which share the same adopted initial $D_{\rm M}$ value, exhibit different final thickness profiles. The final vertical heating parameter of M4 is $H_z=105\,(\mathrm{km\,s^{-1}})^2$, only about 17\% larger than that of M1, whereas its mean relative thickness, $\langle h_z/R_{\rm D}\rangle_R=0.106$, is about 47\% larger. The values of $h_z/R_{\rm D}$ in M4 range from 0.045 to 0.121 across the sampled radii. M4 initially has half the baryonic mass of M1 at unchanged disc scale lengths, and hence half the initial baryonic surface density at a given radius. Its larger relative thickness may therefore reflect weaker absolute vertical confinement, which could allow moderate heating to produce more pronounced thickening. 
 
Taken together, these results show that the vertical evolution of the model sequence varies systematically with the global MOND depth, but is not determined by $D_{\rm M}$ alone. The high-mass, low-$D_{\rm M}$ model M2 experienced the strongest vertical heating, and its $h_z/R_{\rm D}$ clearly exceeded the adopted superthin-disc threshold. Although M3 has the same $D_{\rm M}$ as M2, it retains the baryonic mass of M1 and achieves its lower $D_{\rm M}$ through reduced disc scale lengths. Its weaker vertical heating relative to M2 therefore shows that identical values of $D_{\rm M}$ do not imply identical dynamical evolution. Nevertheless, $h_z/R_{\rm D}$ in the outer disc of M3 approached or exceeded 0.1 at late times, indicating that its compact initial configuration was associated with appreciable outer-disc thickening even though its global vertical heating remained below that of M2. 

Compared with M2 and M3, models M4 and M5, which have higher $D_{\rm M}$ values, maintained thinner stellar discs for most of the simulated evolution. However, when only the late stages of the evolution are considered, $h_z/R_{\rm D}$ in M4 exceeded the threshold of 0.1 over most of the radial range. Within the present model sequence, M5 shows that a diffuse, high-$D_{\rm M}$ disc can retain its overall superthin character despite persistent non-axisymmetric structure in the outer disc. These comparisons indicate that $D_{\rm M}$ is a useful global descriptor of MOND depth, but that the total baryonic mass, disc scale lengths, and associated surface-density distribution must also be considered when assessing the long-term vertical evolution of superthin discs. 

\section{Conclusions}
\label{sec:conclusions} 
In this work, we performed three-dimensional hydrodynamical $N$-body simulations of the well-studied superthin galaxy UGC 7321 within the MOND framework. Our primary aim was to determine whether an observationally constrained superthin baryonic disc could retain its small vertical thickness over $5.0\,\mathrm{Gyr}$ of isolated evolution. We also constructed a sequence of models spanning different global MOND depths to examine how their vertical heating and thickening depend on the baryonic mass distribution. 

The fiducial model developed a prominent bar and a two-armed spiral pattern during its early evolution and subsequently underwent a pronounced bar-buckling episode. Nevertheless, its vertical thickening remained moderate. In this model, the vertical evolution reflected the balance between the growth of vertical random motions associated with non-axisymmetric structures and the confinement supplied by the Milgromian vertical restoring field. After $5.0\,\mathrm{Gyr}$, most of the stellar disc remained below the adopted superthin-disc threshold, $h_z/R_{\rm D}=0.1$, although the outermost region approached this value. The bar and spiral patterns also drove substantial radial angular-momentum redistribution. However, the comparatively weak vertical heating of the outer disc suggests that much of this transport remained within the disc plane rather than being converted into vertical random motion. 

We further used the MOND depth index, $D_{\rm M}$, as a global measure of the extent to which the baryonic distribution occupies the low-acceleration regime, and compared five models spanning three values of $D_{\rm M}$. Within the present model sequence, the lower-$D_{\rm M}$ models generally experienced stronger vertical heating and thickening, whereas the higher-$D_{\rm M}$ models retained thinner stellar discs. However, models with identical values of $D_{\rm M}$ developed different morphologies and heating histories when their total baryonic masses or disc scale lengths differed. The high-mass, low-$D_{\rm M}$ model M2 underwent the strongest vertical heating, while the compact low-$D_{\rm M}$ model M3 experienced weaker global heating but appreciable late-time thickening in its outer disc. The MOND depth index is therefore a useful global descriptor, but it does not uniquely determine the vertical evolution of a superthin disc. The total baryonic mass, disc scale lengths, surface-density distribution, and resulting non-axisymmetric evolution must also be taken into account. Taken together, our results indicate that, under isolated conditions, long-lived superthin stellar discs are dynamically viable within MOND, but their survival depends on the underlying baryonic structure and evolutionary conditions. 

The $D_{\rm M}$ values used in the present comparison are evaluated
from the initial baryonic distributions and are therefore single-epoch structural descriptors rather than conserved dynamical quantities. The baryon-based structural framework introduced by \citet{Eappen2026} provides a useful basis for extending this comparison from initial MOND depths to evolutionary trajectories. The fact that models with the same initial $D_{\rm M}$ can follow different evolutionary paths raises the question of whether the time evolution of the MOND depth index contains information beyond its initial value. With a fixed MOND acceleration scale $a_0$, the index can be evaluated at each time as
\begin{equation}
D_{\rm M}(t)
=
1-\frac{M_{\rm bar}(<r_{\rm M}(t),t)}
        {M_{\rm bar}(t)},
\qquad
r_{\rm M}(t)
=
\sqrt{\frac{G M_{\rm bar}(t)}{a_0}}.
\end{equation}

For an isolated system with approximately conserved baryonic mass, $r_{\rm M}$ remains nearly constant, so changes in $D_{\rm M}(t)$ mainly trace mass redistribution across this radius. Comparisons with $h_z/R_{\rm D}$, $\sigma_z$, and the non-axisymmetric amplitudes could test whether the preceding
history of $D_{\rm M}$ informs subsequent vertical evolution beyond its initial value. 

In a cosmological setting, the index could be followed along a galaxy's assembly history as $D_{\rm M}(z)$, where $z$ denotes cosmological redshift. Such tracks could distinguish systems with similar present-day MOND depths but different histories of baryonic concentration and expansion, potentially helping to explain their different final vertical structures.
Interpreting these tracks would require accounting for changes in baryonic mass and structure through accretion, mass loss, mergers, and environmental effects.
For non-isolated systems, the MOND external field effect must also be considered, since it is not explicitly encoded in this baryon-based index. 

Future work should examine the sensitivity of these results to the adopted effective isothermal gas temperature. The gas component influences the gravitational field through its self-gravity and affects the growth of non-axisymmetric structures through dissipation and pressure support. It can therefore modify the subsequent vertical heating of the stellar disc. In the present isothermal treatment, the adopted effective gas temperature sets the sound speed and pressure support and should not be interpreted as the literal thermodynamic temperature of a multiphase interstellar medium. In their MOND simulations of M33, \citet{Banik2020} found that changing the adopted effective isothermal gas temperature altered the late-time bar and central morphology. A lower value reduced the development of overly strong non-axisymmetric structure and produced a weaker bar that was more consistent with the observed morphology. 

Observations also indicate that the outer \ion{H}{i} disc of UGC 7321 is warped, although the origin of this feature remains uncertain~\citep{Uson2003}. Possible mechanisms include external tidal perturbations, inclined gas accretion, and misalignment between the angular-momentum vectors of the inner and outer discs. Within MOND, the external field effect may provide an additional contribution. Because UGC 7321 is generally regarded as relatively isolated, any relevant external field would more plausibly arise from its large-scale gravitational environment than from a nearby massive companion. Whether such a field is sufficiently strong to produce the observed mild warp remains to be tested through dedicated simulations. 
 
Future simulations could impose static or time-varying external fields with different strengths and orientations on the observationally constrained model. Such calculations would allow the amplitude and radial extent of the outer \ion{H}{i} warp to be quantified and would test whether the external field also modifies the global vertical thickness of the stellar disc. The analysis should also be extended to a larger sample of superthin galaxies to determine which of the trends identified here are generic and which depend on the particular baryonic structure of UGC 7321. 

\begin{acknowledgements}
This work is supported by the National Natural Science Foundation of China under Grant No.~11965016 and by the Natural Science Foundation of Qinghai Province under Grant No.~2022-ZJ-939Q. 
\end{acknowledgements}

\bibliographystyle{bibtex/aa}
\bibliography{bibtex/myBiblio}

\begin{appendix}
\onecolumn
\section{Additional resolution and MOND-depth diagnostics}
\label{app:diagnostics}
\noindent
\begin{minipage}{\textwidth}
    \centering
    \includegraphics[width=1.0\textwidth]{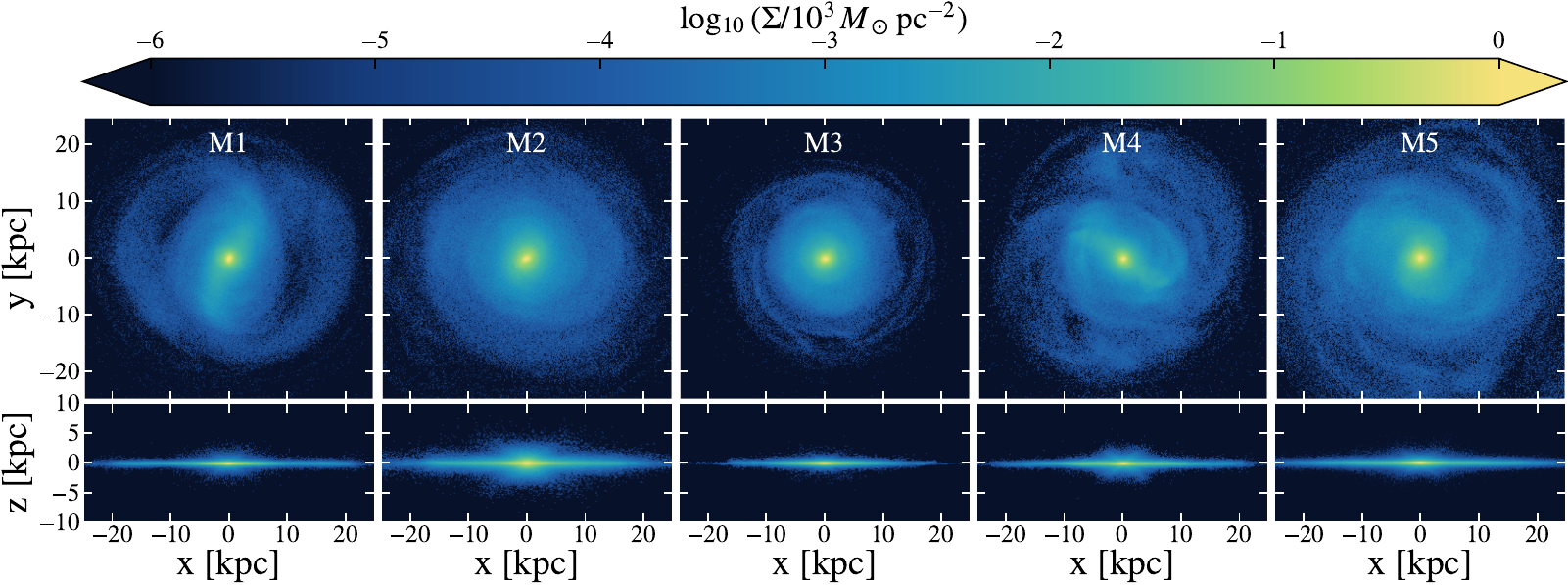}
    \captionof{figure}{Final stellar morphologies of models M1--M5 at $t=5.0\,\mathrm{Gyr}$. From left to right, the columns correspond to models M1--M5. The upper and lower rows show the face-on and edge-on projections of the stellar discs, respectively.}
    \label{fig11}
    \vspace{0.35em}
    \begin{minipage}[t]{0.485\textwidth}
        \centering
        \vspace{0pt}
        \includegraphics[width=0.96\linewidth]{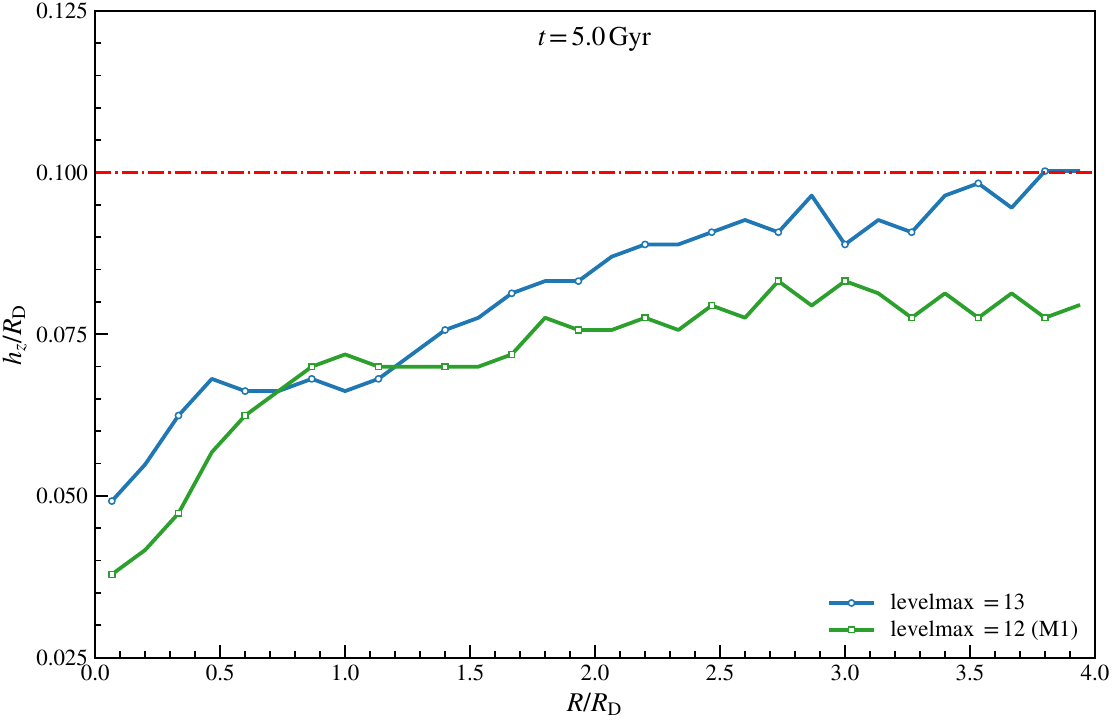}
        \captionof{figure}{Final $h_z/R_{\rm D}$ profiles of the high-resolution fiducial model (levelmax=13; blue circles) and the lower-resolution M1 model (levelmax=12; green squares) over $0\leq R/R_{\rm D}<4$ at $t=5.0\,\mathrm{Gyr}$. The red dash-dotted line marks $h_z/R_{\rm D}=0.1$.}
        \label{fig12}
    \end{minipage}
    \hfill
    \begin{minipage}[t]{0.485\textwidth}
        \centering
        \vspace{0pt}
        \includegraphics[width=0.96\linewidth]{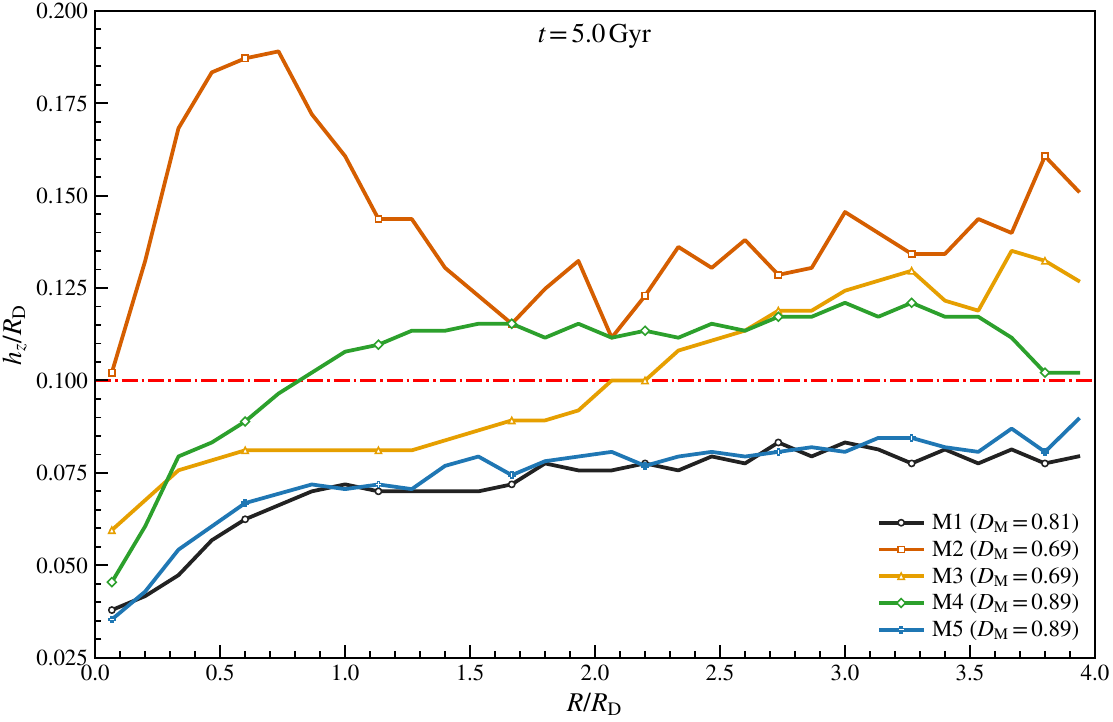}
        \captionof{figure}{Final $h_z/R_{\rm D}$ profiles of models M1--M5 over $0\leq R/R_{\rm D}<4$ at $t=5.0\,\mathrm{Gyr}$. The legend gives the adopted initial $D_{\rm M}$ of each model, and the red dash-dotted line marks $h_z/R_{\rm D}=0.1$.}
        \label{fig13}
    \end{minipage}
    \vspace{0.35em}
    \captionof{table}{Final vertical-structure diagnostics of models M1--M5 at $t=5.0\,\mathrm{Gyr}$.}
    \label{table3}
    \setlength{\tabcolsep}{3.0pt}
    \renewcommand{\arraystretch}{0.95}
    {\small
    \begin{tabular}{lcccc}
        \hline\hline
        Model & $D_{\rm M}$ & $\langle h_z/R_{\rm D}\rangle_R$
        & Minimum and maximum values & $H_z$ \\
         & & & & $[(\mathrm{km\,s^{-1}})^2]$ \\
        \hline
        M2 & 0.69 & 0.142 & $0.102$--$0.189$ & 631.0 \\
        M3 & 0.69 & 0.100 & $0.059$--$0.135$ & 130.0 \\
        M1 & 0.81 & 0.072 & $0.038$--$0.083$ & 90.3 \\
        M4 & 0.89 & 0.106 & $0.045$--$0.121$ & 105.0 \\
        M5 & 0.89 & 0.074 & $0.035$--$0.089$ & 91.6 \\
        \hline
        \hline
    \end{tabular}}
    \tablefoot{Here $D_{\rm M}$ is the initial MOND depth index. The mean, minimum, and maximum values of $h_z/R_{\rm D}$ are calculated over valid equal-width radial bins spanning $0\leq R/R_{\rm D}<4$. The interval between the minimum and maximum values describes radial variation rather than statistical uncertainty. Each model is normalised by its fixed initial $R_{\rm D}$. The vertical heating parameter is $H_z(t)=\sigma_z^2(t)-\sigma_z^2(t_0)$, with $\sigma_z$ measured over $0.5\leq R/R_{\rm D}<4$. Rows are ordered by increasing $D_{\rm M}$ to facilitate comparison of the matched-depth pairs.}
\end{minipage}
\clearpage

\begin{figure}[!htbp]
    \centering
    \includegraphics[
        width=1.0\textwidth,
        height=1.0\textheight,
        keepaspectratio
    ]{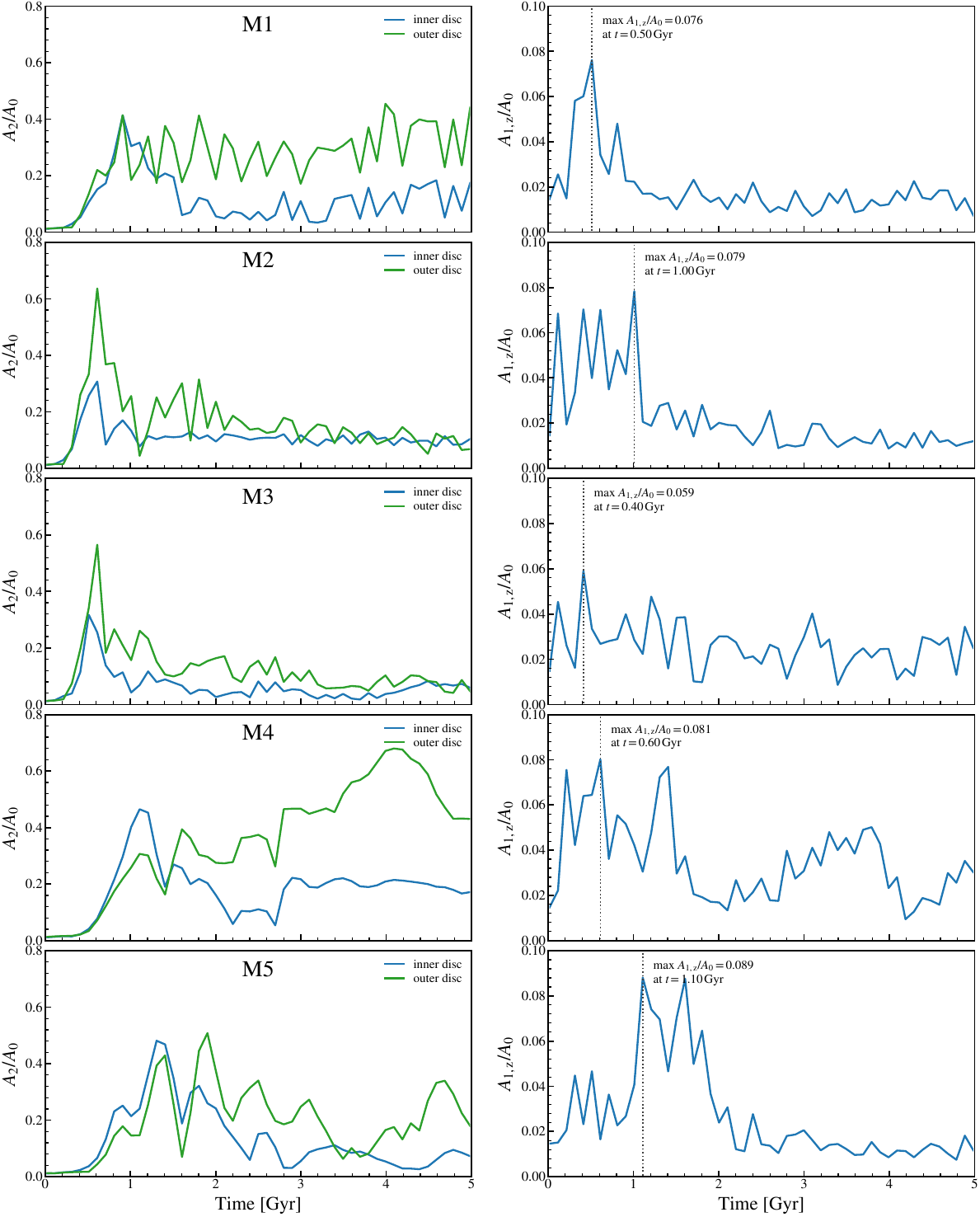}
    \caption{Time evolution of the normalised stellar Fourier amplitudes, $A_2/A_0$ and $A_{1,z}/A_0$, for models M1--M5. From top to bottom, the rows correspond to models M1--M5. The left-hand panels show the normalised $m=2$ amplitude, $A_2/A_0$, measured separately in the inner disc (blue curves) and outer disc (green curves). The right-hand panels show the normalised vertical buckling amplitude, $A_{1,z}/A_0$.}
    \label{fig14}
\end{figure}

\end{appendix}

\end{document}